\documentclass[preprint2]{aastex}
\graphicspath{{./}{../}{../figures/}}
\shortauthors{Lagioia et al.}
\newcommand{\ngc}{NGC\,}
\newcommand{\hst}{\textit{HST}}

\newcommand{\feh}{\rm [Fe/H]}
\newcommand{\afe}{\rm [$\alpha$/Fe]}
\newcommand{\teff}{$T_{\rm eff}$}

\begin{document}

\title{Helium variation in Four Small Magellanic Cloud Globular Clusters\footnote{Accepted for publication on ApJ, 06 December 2018}}
\author{Edoardo P. Lagioia, Antonino P. Milone, Anna F.Marino} 
\affil{Dipartimento di Fisica e Astronomia ``Galileo Galilei''}
\affil{Universit\`{a} di Padova, Vicolo dell'Osservatorio 3, I-35122, Padua, Italy}
\email{edoardo.lagioia@unipd.it}
\and
\author{Aaron Dotter}
\affil{Harvard-Smithsonian Center for Astrophysics, Cambridge, MA 02138, USA}

\begin{abstract}
The multiple stellar populations (MPs) of the $\sim$11-13 Gyr-old Globular
Clusters (GCs) in our Galaxy are characterized by different content of several light
elements. These elements describe well-defined patterns like the C-N and the
Na-O anticorrelations and the He-N and Na-N correlations.

The discovery of the MPs in Magellanic Cloud GCs opened up new paths
for the investigation of chemical anomalies in clusters with different age and
physical properties. 

In this context, we used {\it Hubble Space Telescope} photometry to investigate
the MPs and constrain their chemical composition of four $\sim$6-11 Gyr
extragalactic GCs, namely \ngc121, \ngc339, \ngc416 and Lindsay\,1 in the Small
Magellanic Cloud.

The comparison of the stellar colors with synthetic spectra suggests
that second-population stars of \ngc121, \ngc339, \ngc416 are slightly enhanced
in helium by $\delta Y=0.009\pm0.006, 0.007\pm0.004$ and
$0.010\pm0.003$, respectively, with respect to the first population,
while we find no significant helium variation in Lindsay\,1 $(\delta
Y=0.000\pm0.004)$. Moreover, second-population stars of all the clusters are,
on average, enhanced in nitrogen and depleted in carbon and oxygen, in close
analogy with what we observe in Galactic GCs.

\end{abstract}


\section{Introduction} \label{sec:intro}
The observation of multiple populations (MPs) in the old ($\sim$11-13 Gyr)
Globular Clusters (GCs) of our Galaxy sets a new standard in the definition of
the nature of these objects. Previously thought to be composed of a single,
coeval generation of stars, GCs in fact host distinct groups of stars with
chemical properties characterized by specific patterns defined by the
anti-correlation of the light elements C-N and Na-O \citep{kraft94,gratton12},
the difference in helium content and, in some cases, by metallicity variations
\citep{marino15}.  In this regard, the determination of internal helium
variations among GC stars is of primary importance because it provides compelling
information about the physical processes that led to the formation of MPs
\citep{decressin07a,dercole08,bastian13c,dantona16}; allows to constrain models
of stellar structure and evolution \citep{cassisi17}; ultimately, gives
important clues about the role played by the first stars in the reionization of
the Universe at high-redshift \citep{schaerer11,renzini17}. 

Direct spectroscopic measurements of absolute helium abundance in old GCs, however,
are limited to hot horizontal branch (HB) stars in the effective temperature
(\teff) range 8,000-11,500 K \citep[e.g.][]{villanova09a,marino14a}. In hotter
stars, indeed, the pristine atmospheric chemical composition is altered by
radiative levitation of metals and gravitational settling of helium
\citep{grundahl99,behr03,moehler04}. In addition, for  some GCs, spectroscopic
helium estimates have been derived from chromospheric lines of few stars at the Red
Giant Branch (RGB) tip \citep{pasquini11,dupree11}.

Relative helium variations between MPs in a GC can be estimated, instead, for a large
number stars. An increase in helium content results, in fact, in a larger \teff\ 
hence in a color change, at a given luminosity, for stars along the Main
Sequence (MS) and the RGB. Several works, based on \textit{Hubble Space
Telescope} (\hst) multi-wavelength photometry, have used this property to infer
the difference in helium mass fraction ($\delta Y$) in MPs of about 60 Galactic GCs
with a precision better than 0.01 \citep{milone12c,milone15a,milone18b}.

Helium variations affect also the stellar luminosities and, as a consequence,
the location of the characteristic evolutionary features in a color-magnitude
diagram (CMD). In particular, the brightness of the RGB Bump (RGBB) is directly
related to the helium content of the underlying stellar population
\citep[see][and references therein]{cassisi97c}. This implies that any
difference in luminosity between the RGBBs of distinct populations in a
monometallic (same [M/H]) GC can be used to infer their relative
helium content. Recent analyses have provided the first $\delta Y$ estimates
from the RGBB of MPs for a sample of about 20 Galactic GCs
\citep{lee15,milone15b,lee17,lagioia18,lee18}.

Both helium and light-element variations are crucial observational constraints to test the
validity of any proposed theory of formation of MPs in GCs. The observed chemical
patterns, indeed, suggest that MPs are the result of internal enrichment processes
according to which secondary generations of stars form in an intracluster medium polluted
with material processed and ejected from the pristine stellar generation
\citep[see][and references therein]{renzini15}. Since the formation of the subsequent
stellar generations occurs at the very early stages of cluster life and,
therefore, in different physical conditions, the search
for signature of the presence of MPs in clusters with different age and in
different environment is of primary importance.

While the relative helium abundance of MPs has been investigated in a large
number of Galactic GCs, the helium content of MPs in
extragalactic GCs is still unexplored. In this respect, the GCs in the Magellanic
Clouds represent a valuable statistical sample. These extragalactic systems,
indeed, are close enough to be resolved into single stars and span a wide range
of ages, going from few Myrs to $\sim 11$ Gyr
\citep[e.g.][]{johnson99,bertelli03,mackey04,glatt08b,glatt08a}.

In the last years several works have been dedicated to the study of MPs in the
Magellanic Cloud GCs. While clusters older than $\sim 2$ Gyr exhibit 
light-element variations and multiple RGBs in CMDs obtained with ultraviolet filters
\citep[e.g.][]{muccia09,daless16,nieder17a,nieder17b,martocchia18a}, clusters
younger than $\sim 2$ Gyr do not show any star-to-star abundance variations
\citep{muccia11,muccia14c} but, rather, some peculiar photometric features,
namely split MS and/or extended MS turn-off, which can be interpreted as due to
age spread caused by prolonged star formation episodes
\citep[e.g.][]{mackey07b,mackey08,goud09,milone09,goud11a} or to a difference in
rotational velocity of MS stars
\citep[e.g.][]{dantona15,bastian16,milone16b,milone17b,milone18a}.

In this paper we will exploit multi-band \hst\ photometry to infer, for the
first time, the helium content of MPs in four $\sim$6-11 Gyr-old Small
Magellanic Cloud (SMC) GCs, namely \ngc121, \ngc339, and \ngc416 and Lindsay\,1
\citep{daless16,nieder17a,nieder17b}. The paper
is organized as follows. In Section~\ref{sec:data} we describe the data and the
procedure to derive photometry and astrometry. The multiple populations are
identified in Section~\ref{sec:MPs}, while Sections~\ref{sec:He}
and~\ref{sec:RGBB} are devoted to the determination of the helium abundance from
the colors of RGB stars and from the magnitudes of the RGBBs, respectively.
Summary and discussion are provided in Section~\ref{sec:conclus}.

\section{Observations and data reduction} \label{sec:data}
For our analysis we took advantage of
archival\footnote{\url{https://archive.stsci.edu/}} \hst\ observations of
\ngc121, \ngc339, \ngc416 and  Lindsay\,1. In particular we used images collected
with WFC3/UVIS in the UV bands F336W and F343N and in the optical band F438W,
and with ACS/WFC in the optical bands F555W and F814W. For \ngc121,
the dataset also includes F814W images collected with WFC3/UVIS. A summary of
the observations used in this work is provided in Table~\ref{tab:tab1}.

%%%%%%%%%%% table %%%%%%%%%%
\begin{deluxetable}{cccccc}
\tabletypesize{\small}
\tablewidth{0pt}
\tablecaption{Observation dataset.\label{tab:tab1}}
\tablehead{
\colhead{Cluster} & \colhead{date} & \colhead{camera} & \colhead{filter} & \colhead{No.$\times$exposure time (s)} & \colhead{Proposal ID}}
\startdata
\ngc121    & 21 Jan 2006               &  ACS/WFC  & F555W & $2\times20 + 4\times496$              & 10396 \\
     	   &                           &           & F814W & $2\times10 + 4\times474$              &       \\
     	   & 16 May 2014, 16 Oct 2014  & WFC3/UVIS & F336W & $4\times1061$                         & 13435 \\
     	   &                           &           & F438W & $4\times200$                          &       \\ 
     	   &                           &           & F814W & $2\times100$                          &       \\ 
           & 01 May 2016               &           & F343N & $500 + 800 + 1650$                    & 14069 \\ 
\ngc339    & 28 Nov 2005               &  ACS/WFC  & F555W & $2\times20 + 4\times496$              & 10396 \\
     	   &                           &           & F814W & $2\times10 + 4\times474$              &       \\
     	   & 08 Aug 2016               & WFC3/UVIS & F336W & $700 + 1160 + 1200$                   & 14069 \\
           &                           &           & F343N & $520 + 800 + 1250 + 1650$             &       \\ 
     	   &                           &           & F438W & $120 + 180 + 560  + 660$              &       \\ 
\ngc416    & 22 Nov 2005, 08 Mar 2006  &  ACS/WFC  & F814W & $4\times10 + 4\times474$              & 10396 \\
     	   & 08 Mar 2006               &           & F555W & $2\times20 + 4\times496$              &       \\
     	   & 16 Jun 2016               & WFC3/UVIS & F336W & $700 + 1160 + 1200$                   & 14069 \\
           &                           &           & F343N & $500 + 800 + 1650 + 1655$             &       \\ 
     	   &                           &           & F438W & $120 + 180 + 560  + 660$              &       \\ 
Lindsay\,1 & 11 Jul 2003, 21 Aug 2005  &  ACS/WFC  & F555W & $2\times20 + 480 + 4\times496$        & 9891, 10396 \\
     	   &                           &           & F814W & $2\times10 + 290 + 4\times474$        &       \\
     	   & 19 Jun 2016               & WFC3/UVIS & F336W & $500 + 2\times1200$                   & 14069 \\
           &                           &           & F343N & $500 + 800 + 1650 + 1850$             &       \\ 
     	   &                           &           & F438W & $120 + 2\times460$                    &       \\ 
\enddata
\end{deluxetable}
%%%%%%%%%%%%%%%%%%%%%%%%%%%%%

The photometric analysis performed on all the images and already described in
\citet{milone18a}, is summarized hereafter. We performed the data
reduction on \textit{flt} images that, in the case of UV observations, were
previously corrected for the poor charge-transfer efficiency following the
recipe of \citet{anderson10}. For each image, we computed a $5\times5$ array of
perturbated point-spread functions (PSFs), starting from library empirical PSFs
and, then, adding spatial-variation corrections obtained from unsaturated and
isolated bright stars. To obtain the position and flux of bright stars we
employed \texttt{img2xym} \citep{and_king06}, a computer program suitably
developed for the reduction of \hst\ data, which is also able to identify saturated
stars and accurately determine their magnitude, by taking into
account the amount of flux bled into adjacent pixels
\citep{gilliland04,anderson08,gilliland10}. The position and flux of faint stars
was, instead, obtained with a different program (Anderson et al. in
preparation), that takes into account all the images in which a stellar image is
present.  Specifications about the algorithm employed by this program are
provided in \citet{sabbi16} and \citet{bellini17a}.

We calibrated the instrumental magnitudes to the VEGAMAG system following
the method of \citet{Bedin05}, using the UVIS and WFC encircled energy
distribution and the photometric zero points available at the STScI
website\footnote{\url{http://www.stsci.edu/hst/wfc3/analysis/uvis_zpts/},\\
\url{http://www.stsci.edu/hst/acs/analysis/zeropoints}}.  The position of stars
was also corrected for geometric distortion by using the solution of
\citet{bellini11} and transformed to the Gaia DR1 reference system \citep{gaia16}.

Finally, we selected for our analysis all the stars measured with high
photometric accuracy, by employing the quality indexes provided with the
software, according to the method detailed in \citet{milone09}.

\begin{figure*}
\centering
\includegraphics[angle=270,width=\textwidth]{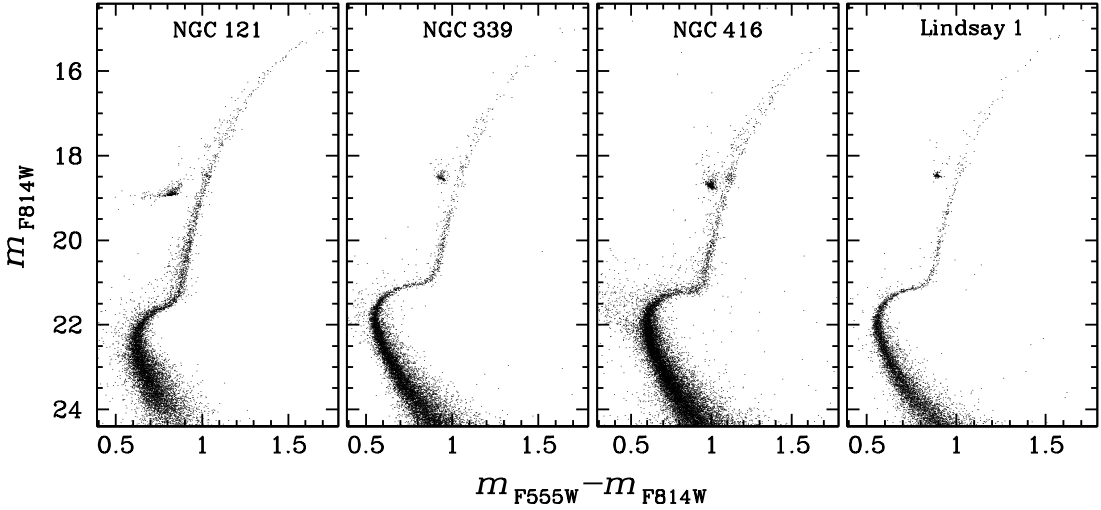} 
\caption{$m_{F814W}$ vs. $m_{F555W}-m_{F814W}$ CMDs  of the four GCs analyzed in
this work. For each cluster only stars within $\sim 50$ arcsec from the cluster center
are displayed.\label{fig:cmds}}
\end{figure*}

The resulting $m_{F814W}$ vs. $m_{F555W}-m_{F814W}$ CMDs of the four GCs are
shown in Figure~\ref{fig:cmds}. In all the cases, we employed the F555W and
F814W magnitudes obtained from ACS observations and displayed all the stars
within 1000 ACS/WFC pixels ($\sim 50$ arcsec) from the cluster
center. Indeed, since our analysis is
aimed at determining the average helium difference of the two main stellar
populations in each GC, we need to select the bulk of cluster members, which are
distributed in the innermost region around the cluster center.  Moreover, the
adoption of the same radius for all the GCs derives from the fact that all the
clusters have approximately the same distance modulus \citep{glatt08b,glatt08a}.
However we verified that the conclusions of our work are not affected by such an
assumption. Therefore, the following analysis is only referred to the sub-sample
of stars included in the aforementioned radial selection.

%%%%%%%%%%%%%%%%%%%%%%%%% FIGURE 1 %%%%%%%%%%%%%%%%%%%%%%%%%%%
%%%%%%%%%%%%%%%%%%%%%%%%%%%%%%%%%%%%%%%%%%%%%%%%%%%%%%%%%%%%

\subsection{Artificial Stars} \label{sec:ASs}
Artificial stars (ASs) were obtained to generate synthetic color-color diagrams
and compare them with observations. The adopted AS test, described in detail in
\citet{anderson08} is briefly summarized here. For each GC we created a list of
$5\times10^5$ ASs, including coordinates and magnitudes, distributed along the
fiducial line of cluster. Then, for each AS in the list, we generated a star
with appropriate position and flux and applied the same procedure and PSF model,
used for real stars, to obtain the photometry of the star.  We also used the
same diagnostics employed for the selection of real stars to assess the
photometric and astrometric quality of the ASs. Finally, we selected only
relatively isolated ASs, with small astrometric and photometric errors, and well
fitted by the adopted PSF model.

\section{Multiple populations along the RGB} \label{sec:MPs}
A glance at the plots in Fig.~\ref{fig:cmds} reveals that every CMD is
compatible with a single stellar population. However, recent works have
shown that, when observed in UV-optical CMDs, all the analyzed GCs display a
split \citep[\ngc121,][]{daless16,nieder17a} or a broadened \citep[\ngc339,
\ngc416, Lindsay\,1][]{nieder17b} RGB, suggesting the presence of multiple
populations.  At variance with pure optical colors indeed, colors obtained from
an appropriate combinations of UV, optical and near-infrared bands are ideal for
the detection of stellar populations characterized by different chemical content
\citep{marino08,milone12b,milone13a,monelli14}. 

For this reason, we devised a procedure for the selection of the MPs along the
RGB of each analyzed cluster, based on the comparison of the distribution of the
observed and artificial stars in the $m_{F336W}-m_{F555W}$ vs
$C_{F343N,F438W,F814W}$ pseudo color-color diagram, with
$C_{F343N,F438W,F814W} = (m_{F343N}-m_{F438W})-(m_{F438W}-m_{F814W})$.

We empirically verified that this diagram maximizes the separation of the RGB
stars with different chemical content. It exploits, indeed, the property of the
narrow UV filter F343N, whose transmission curve is centered on the absorption
band of the NH molecule, and of the F336W and F438W filters, which are similar
to the Johnson broadband filters $U$ and $B$ and are sensitive to the nitrogen
and carbon content, respectively\footnote{The choice of the
$m_{F336W}-m_{F555W}$ vs $C_{F343N,F438W,F814W}$ color-color
diagram to select the two cluster stellar populations, rather than the
$m_{F814W}$ vs $C_{F343N,F438W,F814W}$ pseudo CMD, is also justified by
two additional reasons: the sensitivity of the $m_{F336W}-m_{F555W}$
index to the nitrogen stellar content through the band F336W \citep[see
e.g.][]{milone12b}; the independent information provided by the two color
indices $m_{F336W}-m_{F555W}$ and $C_{F343N,F438W,F814W}$.}. 
We also note in passing that $C_{F343N,F438W,F814W}$ is similar to the color
index $c_{\,UBI}$ for the Johnson-Cousin filter system
\citep{milone12c,monelli13b}.

\begin{figure*}
\centering
\includegraphics[width=\textwidth]{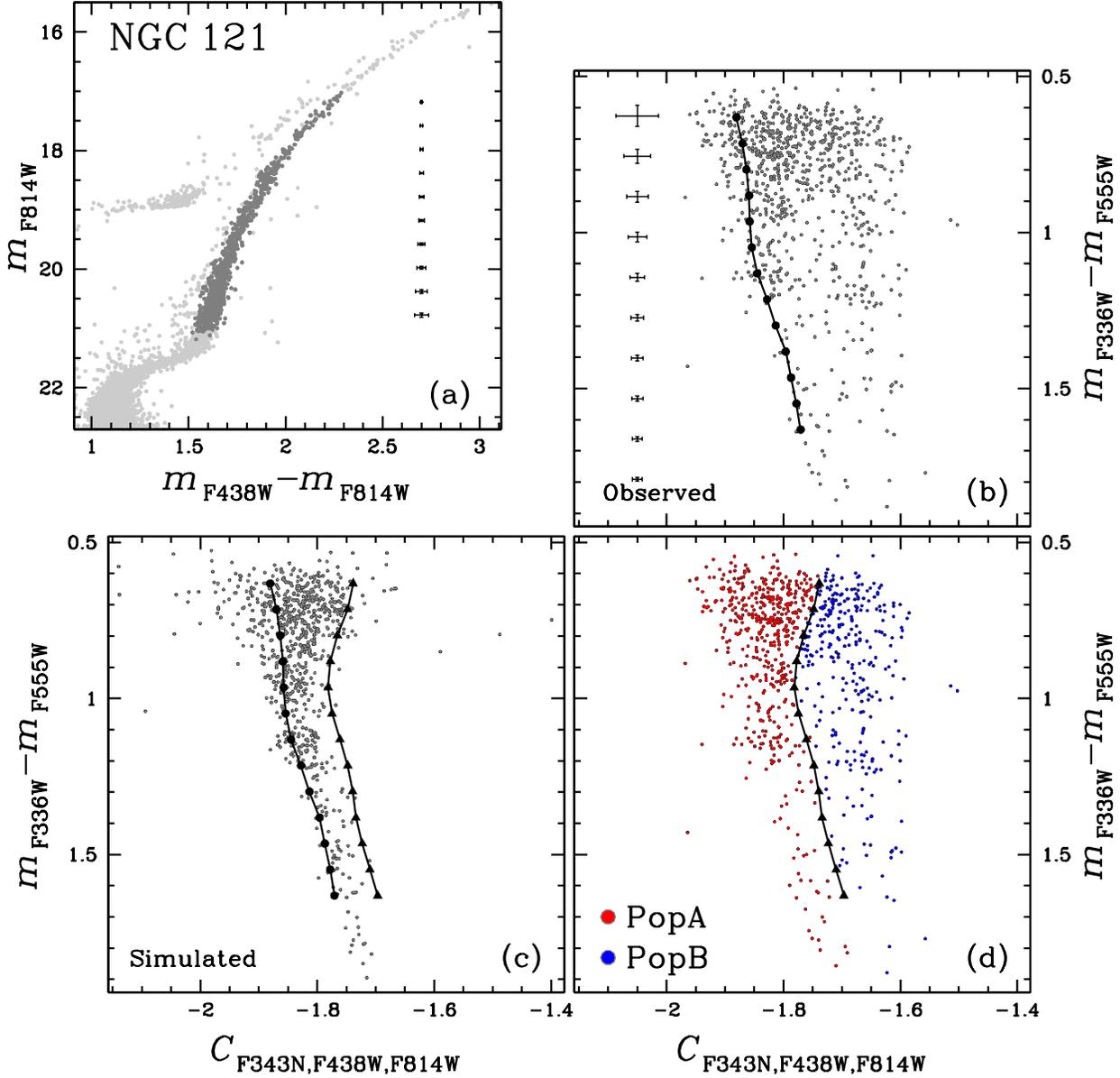}
\caption{Selection of the two main stellar populations of \ngc121.
\textit{Panel (a)}: the cluster RGB members are represented by the dark-gray
points in the $m_{F814W}$ vs. $m_{F438W}-m_{F814W}$ CMD. The error bars
indicate the typical photometric uncertainties of the RGB stars as a function of
the magnitude. \textit{Panel (b)}: $m_{F336W}-m_{F555W}$ vs
$C_{F343N,F438W,F814W}$ pseudo color-color diagram of the observed RGB members.
The black dots, connected by the solid line, mark the 10th percentile of the
$C_{F343N,F438W,F814W}$ pseudo-color distribution of the stars along the
$m_{F336W}-m_{F555W}$ direction. The error bars represent the typical
color uncertainties as a function of the vertical coordinate.
\textit{Panel (c)}: same pseudo color-color diagram as in panel (b) but for the
RGB ASs. The black dots and triangles mark, respectively, the 10th and 98th
percentile of the ASs $C_{F343N,F438W,F814W}$ pseudo-color distribution.
\textit{Panel (d)}: the ASs 98th percentile distribution divides the observed RGB
stars into two groups, named PopA (red points) and PopB (blue points).
\label{fig:mp121}} 
\end{figure*}

An example of the procedure, applied to the case of \ngc121, is given in
Figure~\ref{fig:mp121}, where in panel (a) is shown the $m_{F814W}$ vs.
$m_{F438W}-m_{F814W}$ CMD of the cluster, with the RGB members represented as
dark gray points. The selection of the RGB sample was done by considering the
location of `bona-fide' RGB stars in the $m_{F555W}$ vs.  $m_{F336W}-m_{F555W}$
and in the $m_{F814W}$ vs.  $m_{F438W}-m_{F814W}$ CMD of the cluster. Owing to
low statistics we excluded stars in the proximity of the RGB tip. All the stars
flagged as `bona-fide' RGB stars in both the CMDs were considered RGB members. 

We show, in panel (b), the $m_{F336W}-m_{F555W}$ vs $C_{F343N,F438W,F814W}$
color-color diagram of the selected RGB members. In this diagram the stars seem
to be distributed into two different groups, elongated approximately in the
upper-left – bottom-right direction, for $m_{F336W}-m_{F555W}$ colors redder
than $\sim1$. Similarly, in panel (c), we show the same pseudo color-color
diagram but for the RGB members of the AS sample. Since the ASs were generated,
by construction, using ideal evolutionary sequences represented by the fiducial
lines, their dispersion in the mentioned pseudo color-color diagram is
indicative of the color dispersion that we would ideally observe for a single
stellar population.

With the aim of comparing the color dispersion of the observed and artificial
stars, we divided $m_{F336W}-m_{F555W}$ color range in a regular grid of $C_i$
points $w/3$ mag apart and, for each point, computed the 10th percentile of the
distribution of the $C_{F343N,F438W,F814W}$ values of all the stars in the
interval $(m_{F336W}-m_{F555W})-w/2 < C_i < (m_{F336W}-m_{F555W})+w/2$
\citep{silver86}. Finally, we linearly interpolated the resulting points. We
used a binwidth $w=0.25$ for \ngc121 and \ngc416, $w=0.35$ for \ngc339 and
$w=0.40$ for Lindsay\,1. 

We applied the same procedure for the ASs, but this time we also computed the
points corresponding to the 98th percentile of the distribution of the
$C_{F343N,F438W,F814W}$ values. The $C_{F343N,F438W,F814W}$ values of the ASs
have been then translated so that the line interpolating the 10th-percentile
points of the ASs overlapped that of the observed points.

We have represented the interpolating function of the 10th-percentile points of
the observed stars as a solid line connecting the black dots in panel (b) and
(c), and the interpolating function of the 98th-percentile points function as
the solid line connecting the black triangles in panel (c) and panel (d). As
visible in the latter panel, the 98th-percentile line divides the observed stars
into two groups, that we will conventionally name PopA and PopB. We observe that
PopA stars, represented as red points, attain on average
$C_{F343N,F438W,F814W}$ values bluer than PopB stars, represented as blue
points.

The adopted convention was chosen in analogy to the results obtained from the
comparison of the spectroscopic and photometric features of Red Giants in
Galactic GCs, where Na-poor (O-rich) RGB stars have lower $c_{\,UBI}$ values than
Na-rich (O-poor) RGB stars \citep[e.g.][]{monelli13b,marino17}. For the same reason, we
expect that PopB stars are enhanced in He, N, Na and depleted in C and O, with
respect to PopA stars. In the rest of the paper we will use the
same color-code to represent all the quantities related to PopA and PopB stars.

%%%%%%%%%%%%%%%%%%%%%%%%% FIGURE 2 %%%%%%%%%%%%%%%%%%%%%%%%%%%
%%%%%%%%%%%%%%%%%%%%%%%%%%%%%%%%%%%%%%%%%%%%%%%%%%%%%%%%%%%%

\begin{figure*}
\centering
\includegraphics[angle=270,width=\textwidth]{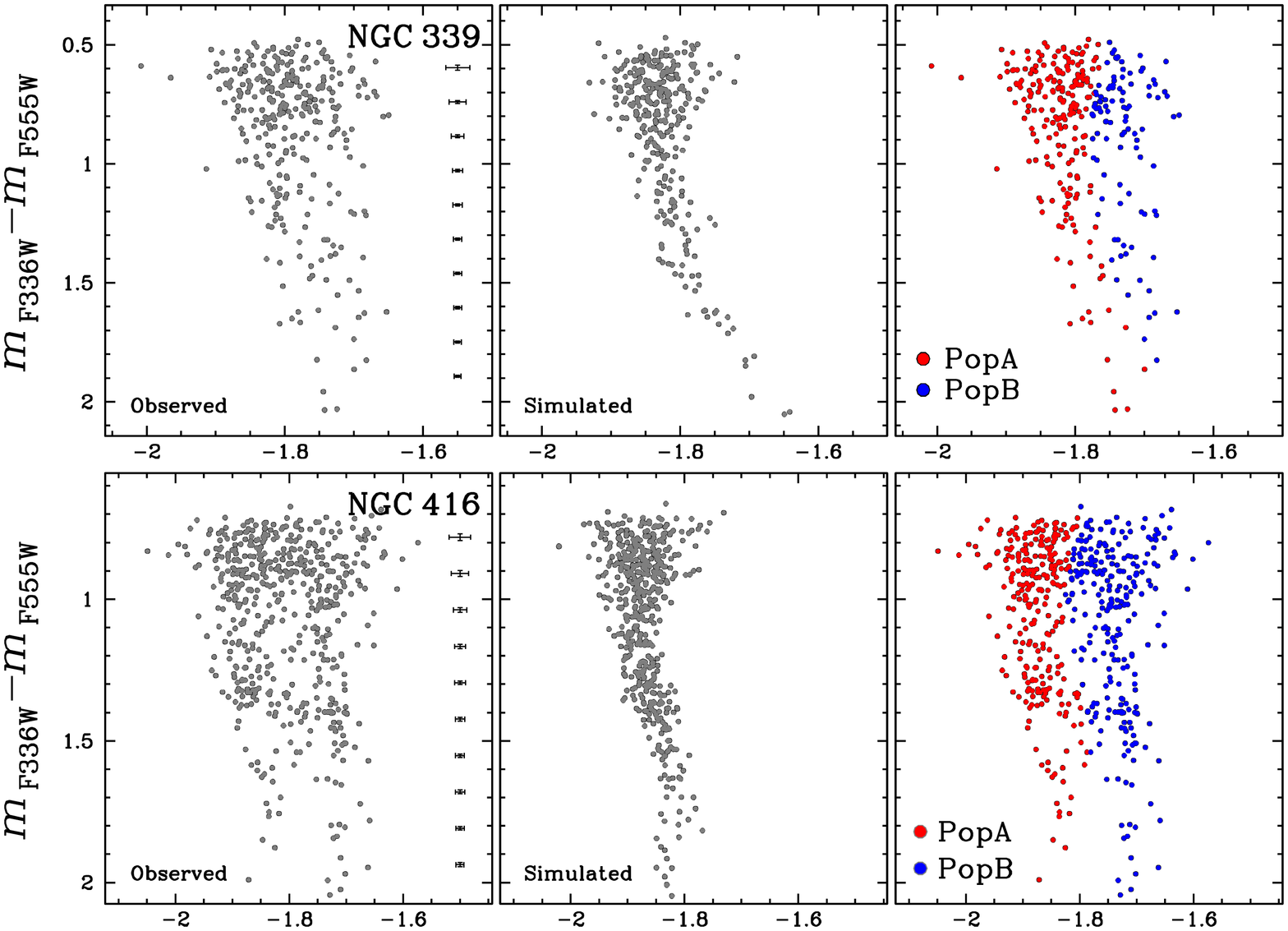}
\includegraphics[angle=270,width=\textwidth]{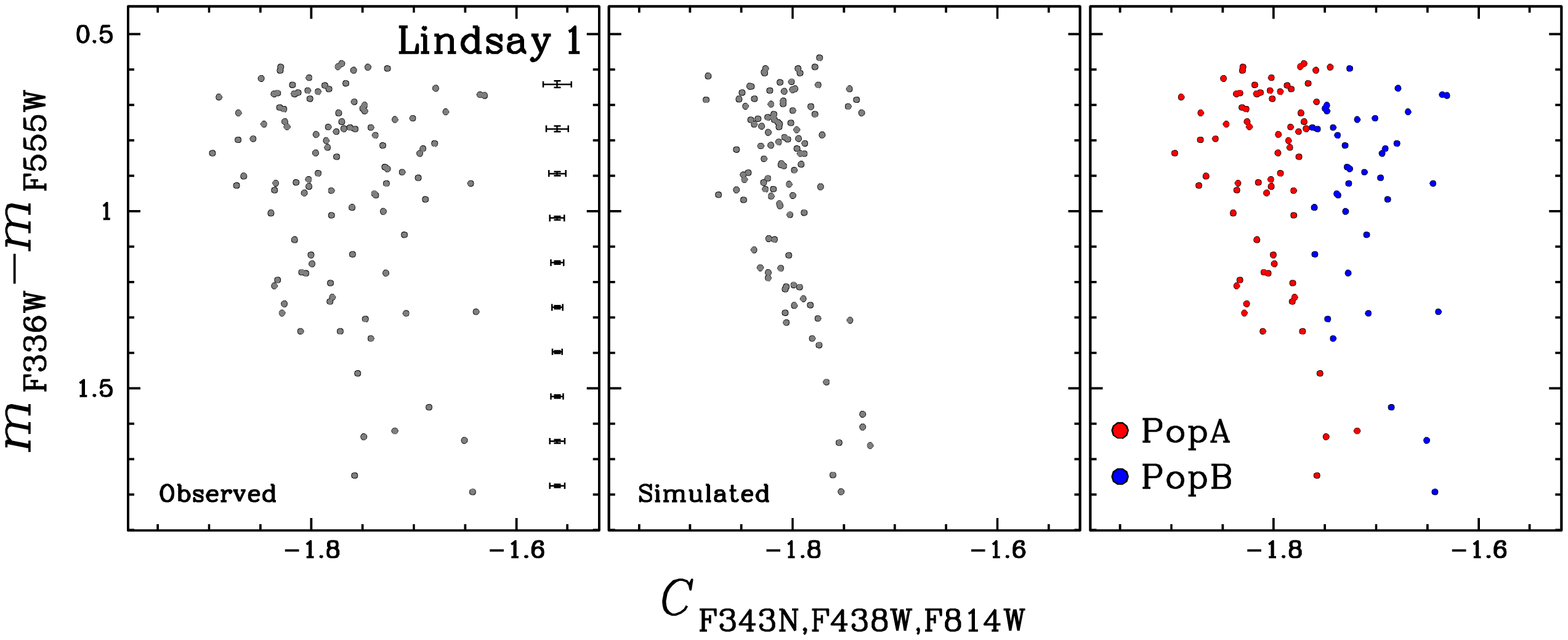}
\caption{\textit{Left panels}: $m_{F336W}-m_{F555W}$ vs $C_{F343N,F438W,F814W}$
pseudo color-color diagram for the observed RGB members of \ngc339, \ngc416 and
Lindsay\,1. The error bars represent the typical color uncertainties as
a function of the $m_{F336W}-m_{F555W}$ color; \textit{Middle panels}:
pseudo color-color diagram for the RGB ASs of each cluster. \textit{Right
panels}: selection of PopA and PopB stars in each cluster.\label{fig:mps}}
\end{figure*}

As already mentioned, we applied the previous method to identify the two main
stellar populations along the RGB of every cluster analyzed in this work.
Figure~\ref{fig:mps} displays, for each labeled GC, the $m_{F336W}-m_{F555W}$ vs
$C_{F343N,F438W,F814W}$ diagram of the observed RGB stars (left panel), the
corresponding diagram for the sample of ASs (middle panel) and the selected PopA
and PopB members (right panel). As in the case of \ngc121, the pseudo
color-color diagram of \ngc416 clearly suggests the presence of two distinct
groups of stars in this cluster \citep[see
also][]{daless16,nieder17a,nieder17b}. On the other side the distribution of
stars in \ngc339 and Lindsay\,1 does not provide the same straightforward
evidence. In all the cases, however, the observed $C_{F343N,F438W,F814W}$
pseudo-color broadening is not compatible with the presence of a single stellar
population.

%%%%%%%%%%%%%%%%%%%%%%%%% FIGURE 3 %%%%%%%%%%%%%%%%%%%%%%%%%%%
%%%%%%%%%%%%%%%%%%%%%%%%%%%%%%%%%%%%%%%%%%%%%%%%%%%%%%%%%%%%

\section{The relative helium content of multiple populations} \label{sec:He}
The average color separation of PopA and PopB stars in a CMD is indicative of
the different chemical content of the two stellar populations \citep{milone15c}.
In particular, since optical bands are mainly sensitive to difference in helium
content through the \teff\ of the stars \citep{lagioia18}, CMDs obtained with a
combinations of optical bands are ideal tools to study helium enrichment in GCs.
In order to quantify the helium abundance variations, $\delta Y$, we used the
procedure already described in \citet{milone12b}, based on the comparison of the
observed color separation of a reference point along each RGB fiducial line with
appropriate theoretical models. 

The main steps of the procedure are summarized hereafter and have been applied
to all the GCs studied in this work. We selected the ACS F814W filter as the
reference band for all the color combinations, $m_{X}-m_{F814W}$, with X =
F336W, F343N, F438W, F555W. Then, we determined the magnitude of the MS turn-off
in the reference band by building the fiducial line of the MS stars, with the
method of the naive-estimator \citep{silver86}: we divided the MS magnitude
range into a grid of $m_{F814W}^i$ points 0.025 mag apart. For each point we
computed the median color, median magnitude and the 68th percentile ($\sigma$)
of the color distribution of all the stars in the interval $m_{F814W}^i-0.125 <
m_{F814W}^i < m_{F814W}^i+0.125$. After performing, in each bin, a
sigma-clipping rejection of all the stars with a distance from the median color
greater than $\sigma$, we computed a new median color, magnitude and $\sigma$.
We smoothed the resulting points with a boxcar average function and, finally, we
linearly interpolated the new points along the MS magnitude interval. The point
with the bluest color along the fiducial line was taken as the MS turn-off. In
similar fashion, we built the fiducial line of the PopA and PopB stars in all
the color combinations. In this case the analyzed magnitude range was divided
into a grid of $m_{F814W}^i$ points $0.5/w$, with $w=1$ for \ngc339, and $w=2$
for \ngc121, \ngc416 and Lindsay\,1. 

We note in passing that in the case of \ngc121, for which observation in the
WFC3 F814W filter are also available, the choice of the latter as the reference
band would not affect the conclusion of the present analysis. We indeed verified
that the average difference between the ACS and WFC3 F814W magnitudes is
consistent with zero for both PopA and PopB stars.

\begin{figure*}
\centering
\includegraphics[width=\textwidth]{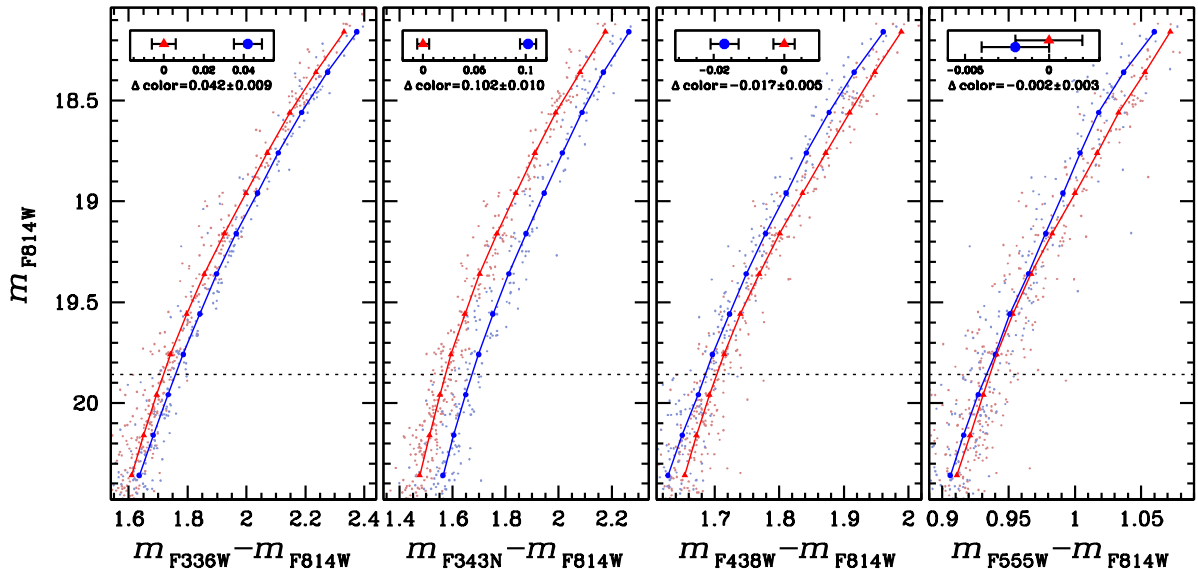}
\caption{Fiducial line of the PopA (light red points) and PopB
(light blue points) stars of \ngc121 in the $m_{F814W}$
vs. $m_{X}-m_{F814W}$ diagrams, with x = F336W, F343N, F438W, F555W. In each
panel, the black dotted line is located at $m_{F814W}^{CUT} = m_{F814W}^{TO} -
2.5$ and the difference between the color of the two fiducials at
$m_{F814W}^{CUT}$, normalized to the PopA fiducial color, is plotted in the
inset. The corresponding color difference is also indicated.
\label{fig:fids121}}
\end{figure*}

As an example, the fiducial lines of the PopA and PopB stars
(represented as light red and light blue points, respectively) of \ngc121 are
shown in Figure~\ref{fig:fids121}, where each panel displays a specific color
combination. The dots along each fiducial line mark the points $m_{F814W}^j =
m_{F814W}^{TO}-(2+0.2 \cdot j)$ with $j=0,1,2,\dots 11$, while the horizontal dotted
line marks the reference magnitude $m_{F814W}^{CUT} = m_{F814W}^{TO}-2.5$. In
each panel, the color difference relative to the color of the PopA fiducial
point, at the reference magnitude, is plotted in the inset. The value of the
color separation, $\Delta$color$ = m_{X}-m_{F814W}$, is also reported with the
associated error, obtained as the sum in quadrature of the standard error of the
color of the PopA and PopB fiducial points, represented by the corresponding
error bar in the plot.

%%%%%%%%%%%%%%%%%%%%%%%%% FIGURE 4 %%%%%%%%%%%%%%%%%%%%%%%%%%%
%%%%%%%%%%%%%%%%%%%%%%%%%%%%%%%%%%%%%%%%%%%%%%%%%%%%%%%%%%%%

\begin{figure*}
\centering
\includegraphics[width=0.7\textwidth]{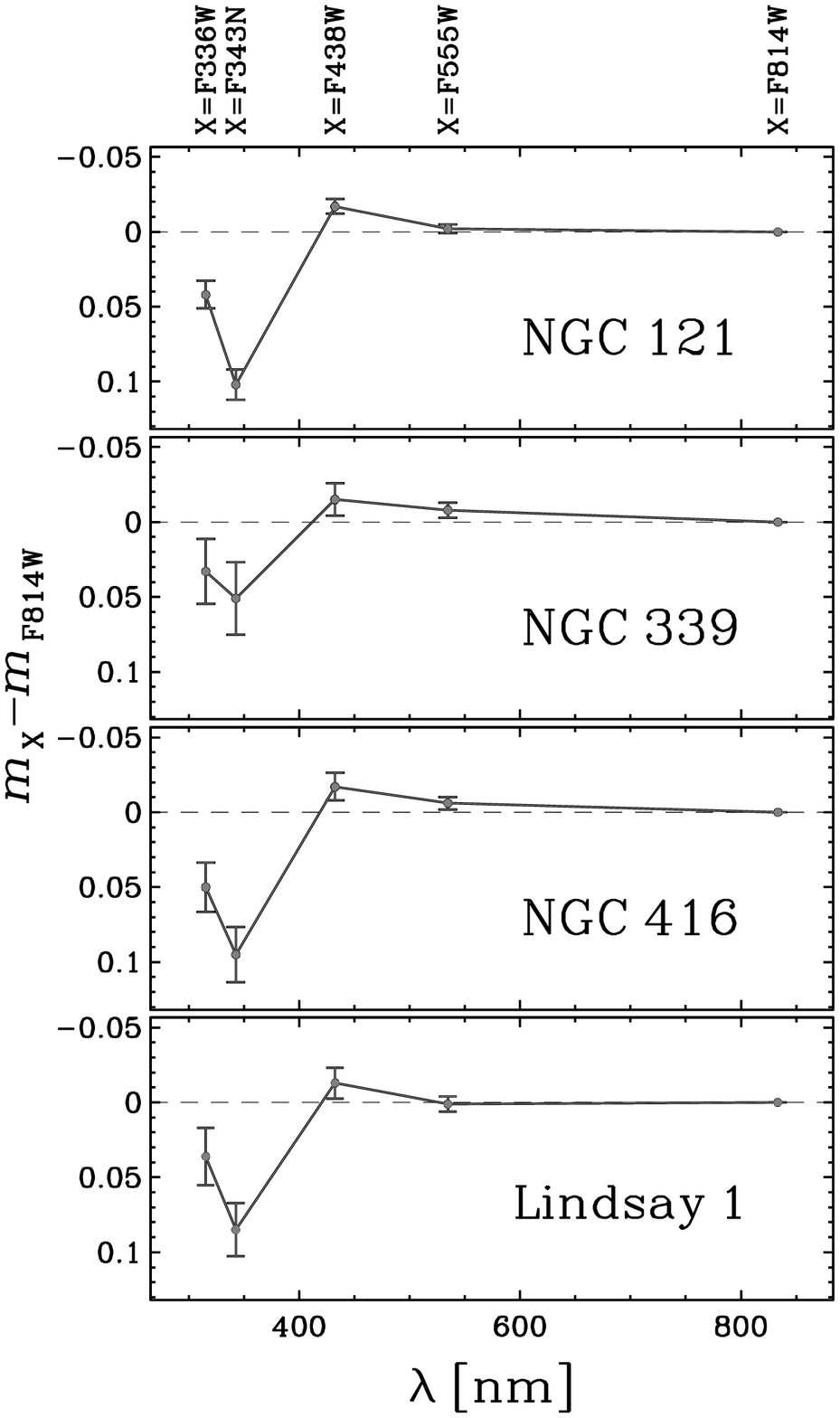}
\caption{Color separation, $m_{X}-m_{F814W}$, with X = F336W, F343N, F438W,
F555W, F814W, of the PopB and PopA fiducial lines at the reference magnitude for
the four GCs studied in this work. The error bars associated to each point
represent the standard error of the color difference.\label{fig:deltacol}}
\end{figure*}

The four panels in Figure~\ref{fig:deltacol} show, for each cluster, the trend
of the color separation as a function of the central wavelength of the
five filters used in this work. By construction the color
separation is zero in F814W band. The error bar associated to each point has
been obtained as the sum in quadrature of the standard errors of the color of
the corresponding PopA and PopB fiducial line points. We observe an overall
comparable trend for all the clusters, with differences attaining positive
values in the UV-optical colors $m_{F336W}-m_{F814W}$ and $m_{F343N}-m_{F814W}$
and negative values in the optical colors $m_{F438W}-m_{F814W}$ and
$m_{F555W}-m_{F814W}$, with the exception of Lindsay\,1 where the difference in
$m_{F555W}-m_{F814W}$ is slightly positive but compatible with zero within the
errors. We also notice a large variation of the color difference values in
correspondence of the narrow bandwidth filter $F343N$ , ranging from 0.051 mag
for \ngc339 to 0.102 for \ngc121. 

As already shown in \citet{milone12b}, the observed trends derive from the
combined effect of the spectral response of the adopted filters and the chemical
properties of the stellar populations. GC stars with primordial composition
(C-rich, N-poor, O-rich) are redder than stars with enhanced composition
(C-poor, N-rich, O-poor) in pure optical colors. Conversely, the former are
bluer than the latter in the $m_{F336W}-m_{F814W}$ or $m_{F343N}-m_{F814W}$
UV-optical colors, because the transmission curve of both the F336W and F343N
filters encompasses the NH absorption band in the typical GC RGB stars
\citep{nieder17a}. In particular the drop of stellar flux is enhanced in the
narrowband filter F343N.

%%%%%%%%%%%%%%%%%%%%%%%%% FIGURE 5 %%%%%%%%%%%%%%%%%%%%%%%%%%%
%%%%%%%%%%%%%%%%%%%%%%%%%%%%%%%%%%%%%%%%%%%%%%%%%%%%%%%%%%%%

\subsection{Synthetic spectra} \label{sec:synsp}
To estimate the relative abundance of C, N, O and He for the two main
populations of each cluster, we compared the observed colors of the PopA and
PopB RGB fiducial lines with the colors derived from synthetic spectra, in close
analogy with what we did in previous papers \citep[e.g.][]{milone12c,milone15a}.

First, we selected four reference magnitude values, $m_{F814W}^l$, where $l =
1,2,3,4$, such that $m^1_{F814W} = m_{F814W}^{TO} -1.5$\footnote{In the case of
\ngc121 we used $m^1_{F814W} = m_{F814W}^{TO} -2.0$} and two adjacent points
differ by $-0.5$ mag. Then, we estimated the \teff\ and gravity associated to
each $m_{F814W}^l$ PopA point, by using isochrones with primordial helium
content, $Y=0.245 + 1.5\cdot Z$, taken from the Dartmouth Stellar Evolution
Database \citep{dotter07,dotter08}. The parameters of the best-fit isochrone
that we found for each cluster are listed in Table~\ref{tab:tab2}. For each
model, we assumed the alpha-element abundance provided in
\citet{glatt08b,glatt08a}.  The metallicity of the selected model was obtained
by best-fitting the cluster RGB. The age was chosen by selecting the central
value between the two models fitting the upper and lower envelope of the upper
MS and Sub-Giant Branch region of the CMD. The typical errors affecting our
determinations are $\pm0.5$ Gyr and $\pm0.1$ dex. The reported values are
consistent with those found by Glatt and collaborators.

We used these parameters to simulate the corresponding stellar spectrum
(hereafter reference spectrum) in the wavelength range between 2,500 and 10,000
\AA. We assumed for the reference spectrum chemical abundances [C/Fe]$=-0.1$,
$[N/Fe]=0.0$ and [O/Fe]$=$\afe.

Then, we computed a grid of synthetic spectra (comparison spectra) with the same
metallicity as the reference spectrum but with Y ranging from the primordial
value to 0.28 in steps of 0.001, [C/Fe] from $-0.1$ to $-0.4$ dex, [N/Fe] from
0.0 to 0.8 dex, and [O/Fe] from the reference alpha abundance values,
$[\alpha/Fe]_{ref}$, listed in Table~\ref{tab:tab2} to
$[\alpha/Fe]_{ref}-0.5$, all in steps of 0.05 dex. Since helium content
significantly affects the structure of a star, hence its temperature and gravity
\citep[see][for discussion]{dotter15}, we assumed for the comparison spectra the
atmospheric parameters derived from to the corresponding helium-enhanced
isochrone. Also, we assumed constant [Fe/H] and overall C$+$N$+$O
content (hence, the effect of metallicty on the stellar structure is the same
for both PopA and PopB stars), in close analogy with what done in previous
papers
\citep[e.g.][]{dotter15,lagioia18,lee18,milone13a,milone15c,milone18b}.
This assumption is supported by the fact that, so far, all the analyzed GCs,
including \ngc121, do not show any evidence of metallicity and C$+$N$+$O variations.
As an example, the cluster SGB morphology is consistent with mono-metallic stellar
populations, as it does not show any evidence for splits and/or spreads, which
are signatures of variation in the overall metallicity
\citep[e.g.][]{yong09,yong14b,marino09,marino12b}.

The spectra, obtained with ATLAS12 and SYNTHE codes
\citep{castelli05,kurucz05,sbordone07}, have been integrated over the
transmission curves of the UVIS/WFC3 and WFC/ACS filters used in this paper to
derive the synthetic colors. The color differences between each comparison
spectrum and the reference spectrum are compared to the observed color
differences between PopB and PopA stars. We assumed, for the abundances of PopB
stars, the content of C, N, O and He of the comparison spectrum that provides
the best fit with the observations.

\begin{figure*}
\centering
\includegraphics[width=0.7\textwidth]{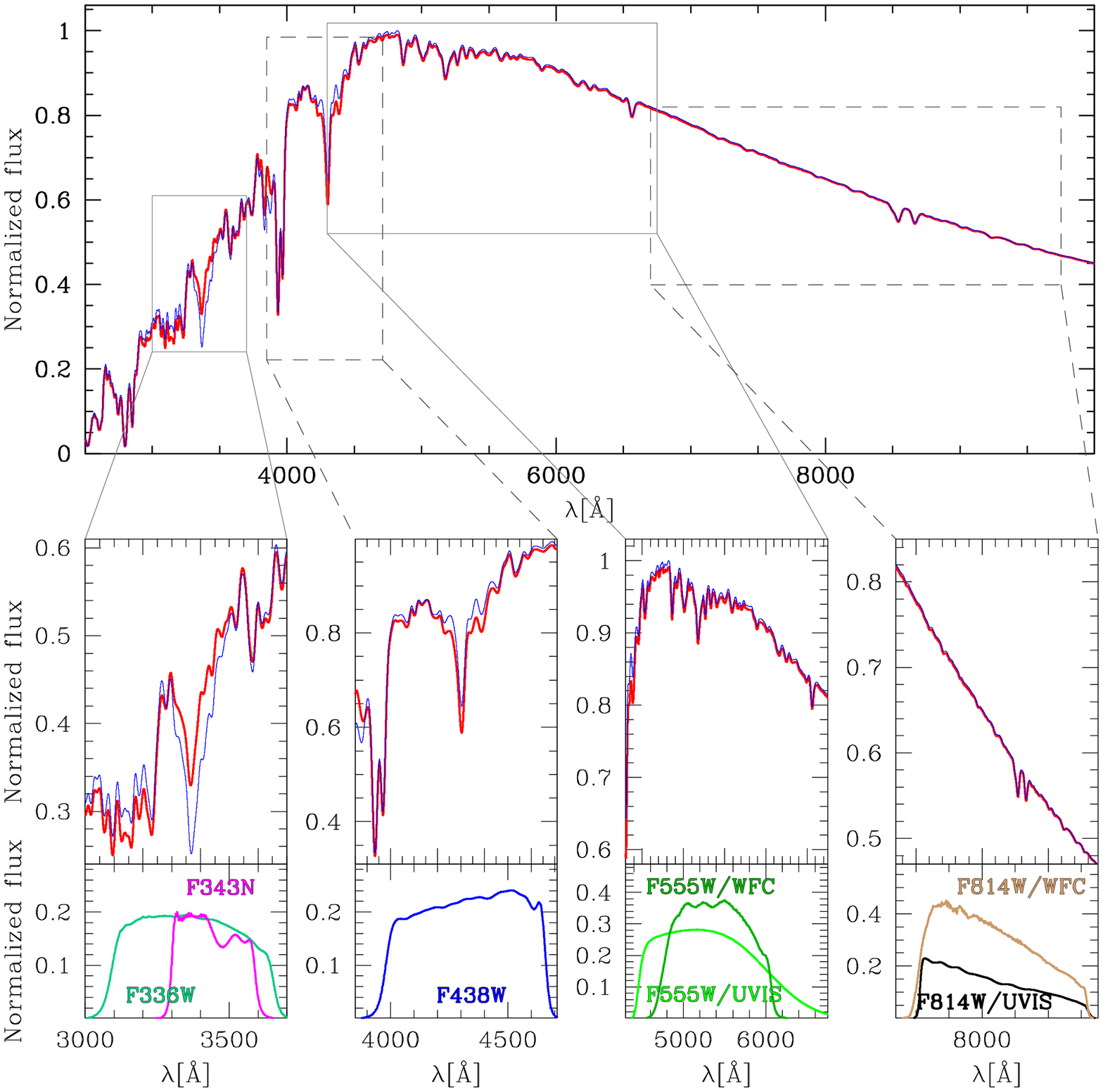}
\includegraphics[width=0.6\textwidth]{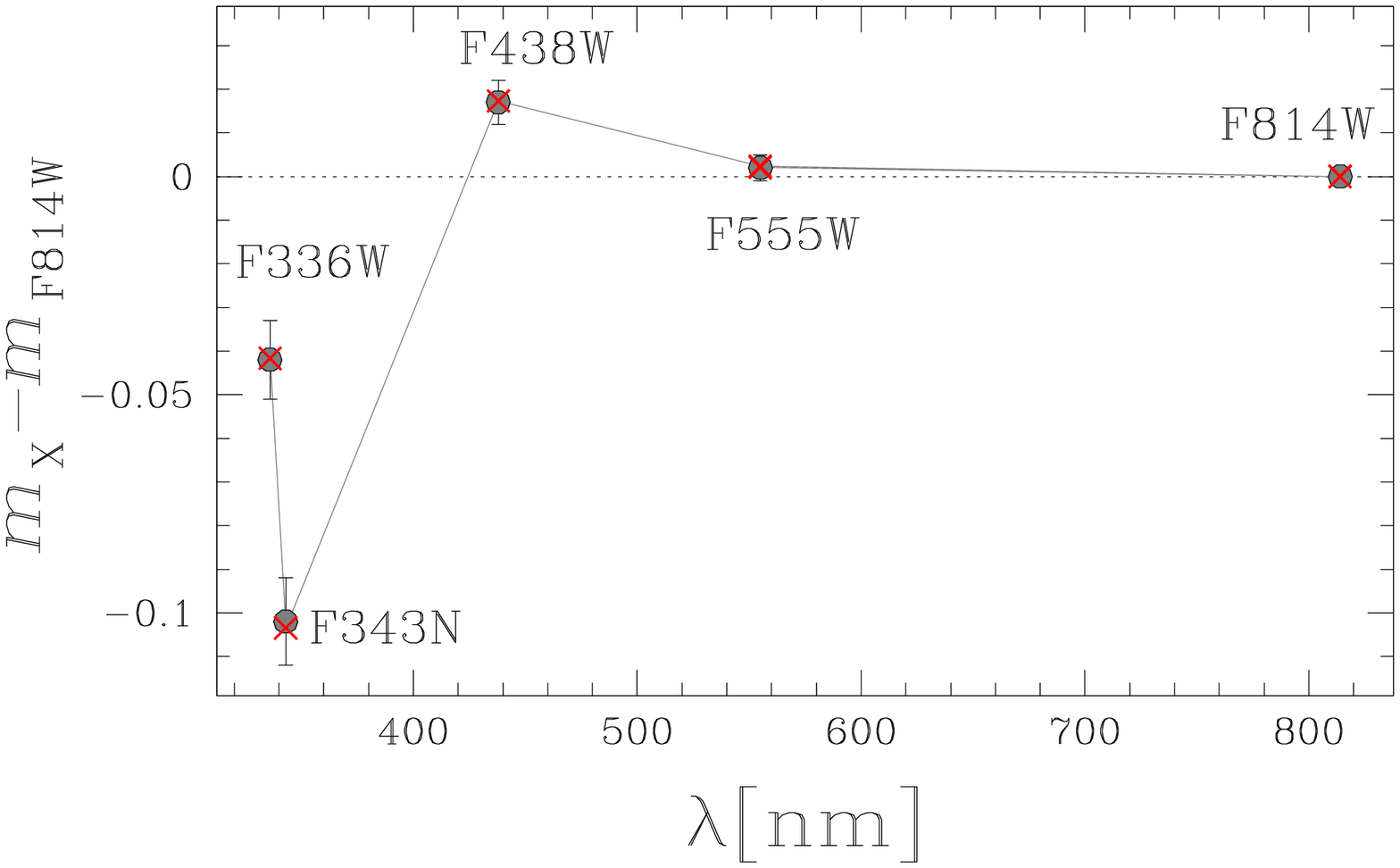}
\caption{\textit{Top panel}: reference spectrum (red; \teff$=5260$ K,
log\,g$=2.90$) and comparison spectrum (blue; \teff$=5274$ K, log\,g$=2.90$) for
the PopA and PopB stars of \ngc121. \textit{Middle panels}: portion of the two
spectra in the regions corresponding to the transmission curves of the filters
used in this paper. \textit{Bottom panel}: comparison between the color
difference of the observed fiducial lines at the reference magnitude (gray dots)
with the same quantity derived from synthetic spectra (red crosses).
\label{fig:synt}}
\end{figure*}

Results are illustrated in Figure~\ref{fig:synt} for \ngc121.  In the top panel
we compare the reference spectrum and the comparison spectrum that best
reproduces the relative colors of PopA and PopB stars with $m_{F814W}=19.864$
(2.5 mag brighter than $m_{F814W}^{TO}$).  The insets are zooms of the two
spectra in the spectral regions of the F336W, F343N, F438W, F555W and F814W
filters, while in the bottom panel we compare the observed color differences
with those derived from the synthetic spectra.

It is important to notice that, since our analysis is based on the
measurement of the relative color of fiducial points, the typical uncertainties
affecting the determination of the best-fit isochrone parameters have a
negligible impact on the derivation of the chemical variations between the two
populations of each cluster. We verified that, for each cluster, a difference of
$\pm0.5$ Gyr in age and $\pm0.1$ dex in metallicity of the adopted models
results in an average difference on the estimate of helium variation equal to
$\langle \Delta(\delta Y)\rangle \lesssim 0.0001$, which in negligible for our
purposes.

%%%%%%%%%%%%%%%%%%%%%%%%% FIGURE 6 %%%%%%%%%%%%%%%%%%%%%%%%%%%
%%%%%%%%%%%%%%%%%%%%%%%%%%%%%%%%%%%%%%%%%%%%%%%%%%%%%%%%%%%%

In Tab.~\ref{tab:tab2} we have reported, for each cluster, the parameters of
the best-fit isochrone (age, \feh, \afe) as well as the relative content of C,
N, O and Y ($\delta$[C/Fe], $\delta$[N/Fe], $\delta$[O/Fe], $\delta Y$) between
the best-fit and the reference synthetic spectrum.  We notice that for
\ngc121, \ngc339 and \ngc416, PopB stars are more helium-rich than PopA stars of
$\sim$~0.01 in mass fraction, while both the stellar populations of Lindsay\,1
have the same helium abundance within the errors.

%%%% table %%%%
\begin{deluxetable}{lccccccccr@{$\,\pm\,$}l}
\tablecaption{Parameters of the best-fit isochrones (columns 2-6) and chemical abundance variations (columns 7-10) between PopA and PopB stars.\label{tab:tab2}}
\tablehead{
 \colhead{Cluster} & \colhead{$\mu_0$} & \colhead{E(B$-$V)} & \colhead{age} & \colhead{\feh} & \colhead{\afe} & \colhead{$\delta$[C/Fe]} & \colhead{$\delta$[N/Fe]} & \colhead{$\delta$[O/Fe]} & \multicolumn{2}{c}{$\delta Y$} \\
   & \colhead{(mag)} & \colhead{(mag)} & \colhead{(Gyr)} & \colhead{(dex)} & \colhead{(dex)} & \colhead{(dex)} & \colhead{(dex)} & \colhead{(dex)} & \multicolumn2c{} 
	}
\startdata
\ngc121    & 18.93 & 0.04 & 10.5 & -1.30 &  0.20 & -0.20 & 0.65 & -0.25 & 0.009&0.006 \\   
\ngc339    & 18.85 & 0.06 &  6.5 & -1.12 &  0.00 & -0.10 & 0.50 &  0.00 & 0.007&0.004 \\   
\ngc416    & 18.96 & 0.09 &  6.0 & -0.96 &  0.00 & -0.15 & 0.60 &  0.00 & 0.010&0.003 \\   
Lindsay\,1 & 18.80 & 0.04 &  8.0 & -1.14 &  0.00 & -0.20 & 0.45 & -0.10 & 0.000&0.004 \\   
\enddata
%\tablenotetext{a}{\citealp{glatt08a,glatt08b}.}
%\tablenotetext{b}{\citealp{daless16}.}
\end{deluxetable}
%%%%%%%%%%%%%

\begin{figure*}
\centering
\includegraphics[width=0.7\textwidth]{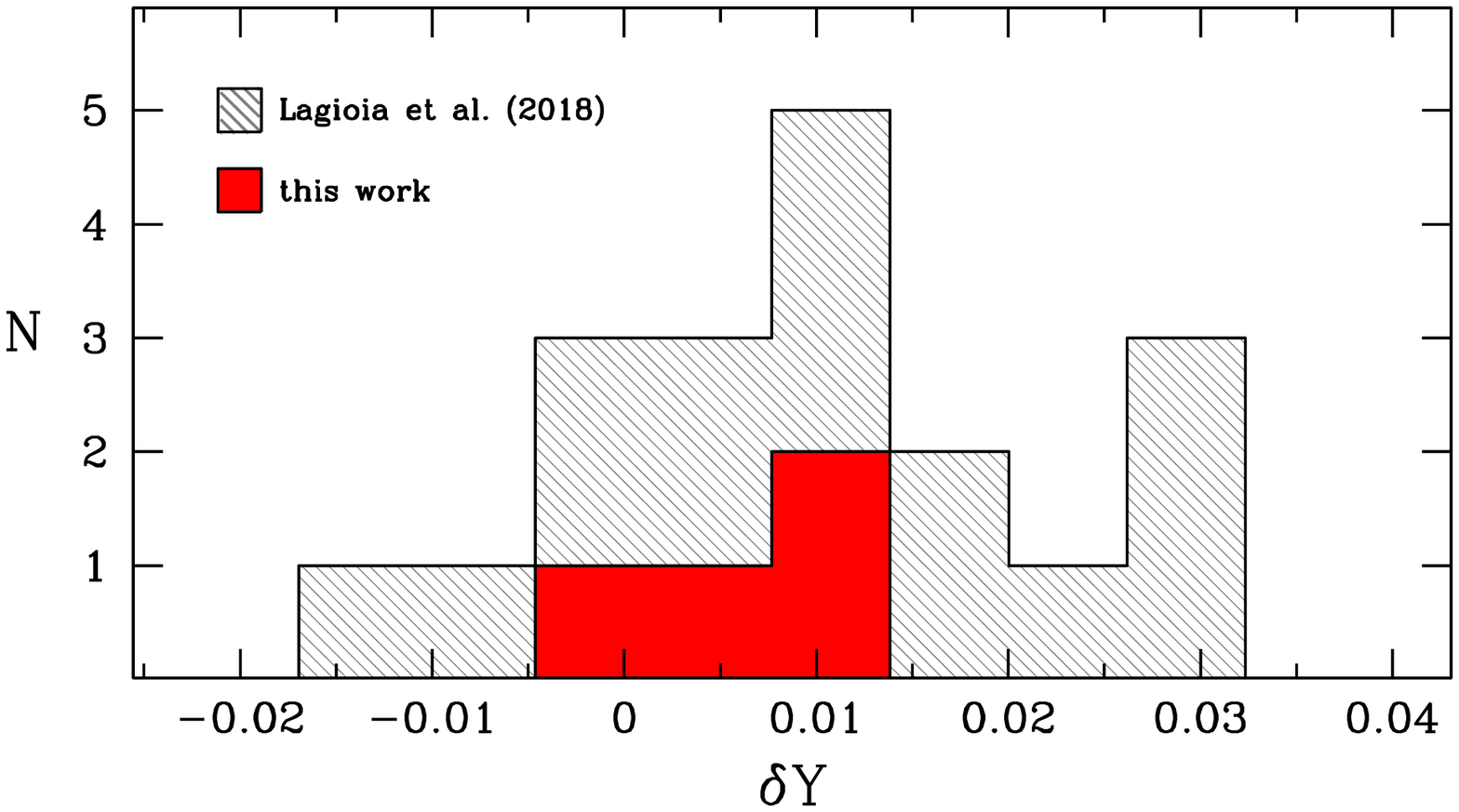}
\caption{Comparison of the distribution of $\delta Y$
values of the four SMC cluster analyzed in this paper (red histogram) with those
found from the analysis of the RGBBs of the Galactic GCs (gray-shaded histogram)
from \citet{lagioia18}.\label{fig:dy}}
\end{figure*}

In a recent paper, we identified the two main stellar populations, namely `1G'
and `2G', in a large sample of Galactic GCs \citep{milone17a} and inferred the
average helium difference between 2G and 1G \citep{lagioia18}. In an attempt to
compare the variation of helium of Milky Way and SMC clusters we overplotted, in
Figure~\ref{fig:dy}, the distribution of the $\delta Y$ values in
Tab.~\ref{tab:tab2} (red histogram) to that obtained from the analysis of the
RGBB by \citet[][see their Fig.10, grey-shaded histogram]{lagioia18}. We see
that the helium-enrichment in the four extragalactic GCs is compatible with that
observed for most Galactic GCs ($ 0.00 \lesssim \delta Y \lesssim +0.01$).
Unfortunately, the selection of the two main stellar populations in the Galactic
GCs is based on photometric diagrams \citep{milone17a} and criteria different
from those adopted in this paper. For this reason, we point out that this
comparison is only indicative since the two groups of stars defined for each
cluster in this work, namely PopA and PopB, do not necessarily correspond to the
stellar populations `1G' and `2G' of the Galactic GCs.

%%%%%%%%%%%%%%%%%%%%%%%%% FIGURE 7 %%%%%%%%%%%%%%%%%%%%%%%%%%%
%%%%%%%%%%%%%%%%%%%%%%%%%%%%%%%%%%%%%%%%%%%%%%%%%%%%%%%%%%%%

\section{Helium estimate from the RGB Bump} \label{sec:RGBB}
One of the most important evolutionary features of stellar populations older
than $\sim 1$ Gyr is the RGBB. It appears as a concentration of stars
along the RGB in CMDs or, equivalently, as a prominent peak in the LF of RGB
stars. The RGBB is produced by a temporary drop of the
luminosity of red giant stars, during their ascent of the RGB, due to the
increase in the gas opacity at the point in which the expanding hydrogen shell
approaches the chemical discontinuity left behind by the deepest penetration of
the convective envelope. Then, once the shell reaches the discontinuity, the
stellar luminosity starts again to increase monotonically
\citep{sweigart90,cassisi16}. A direct consequence of such a mechanism is that
the observational properties of the RGBB are directly related to the physical
and chemical properties of the stars \citep{bono01,nataf14}. For instance,
variations in the chemical content of MPs in a GC will
result in a brightness difference of the corresponding RGBBs. In particular, in
visual bands this phenomenon is mostly connected to variation in helium
abundance \citep{milone15c,lagioia18}.  

As shown in the recent survey of \citet{lagioia18} on the RGBBs of multiple
populations in Galactic GCs, the helium difference between the two main
population of a cluster can be obtained with a method
consisting of three main steps: the construction of the RGB LF for the
measurement of the magnitude difference of the RGBB of the two populations in optical
bands; the use of synthetic stars for the evaluation of the statistical
significance of the peak detections; the comparison of observations with
simulated CMDs obtained from appropriate theoretical models. We decided to apply
this procedure for the estimate of the relative helium abundance of MPs in
\ngc121, that is the only cluster among those analyzed in this paper for which
the RGBB of PopA and PopB stars is visible. Indeed, owing to the low
number of stars in the RGBB region, it was not possible to unambiguously
determine the location of both the PopA and PopB RGBB for the cluster \ngc339 and
Lindsay\,1. Similarly, in the case of \ngc416, it was not possible to determine
the RGBB location of PopB stars due to their flat distribution in the bump
region.

\subsection{The observed luminosity functions}
\begin{figure*}
\centering
\includegraphics[width=0.8\textwidth]{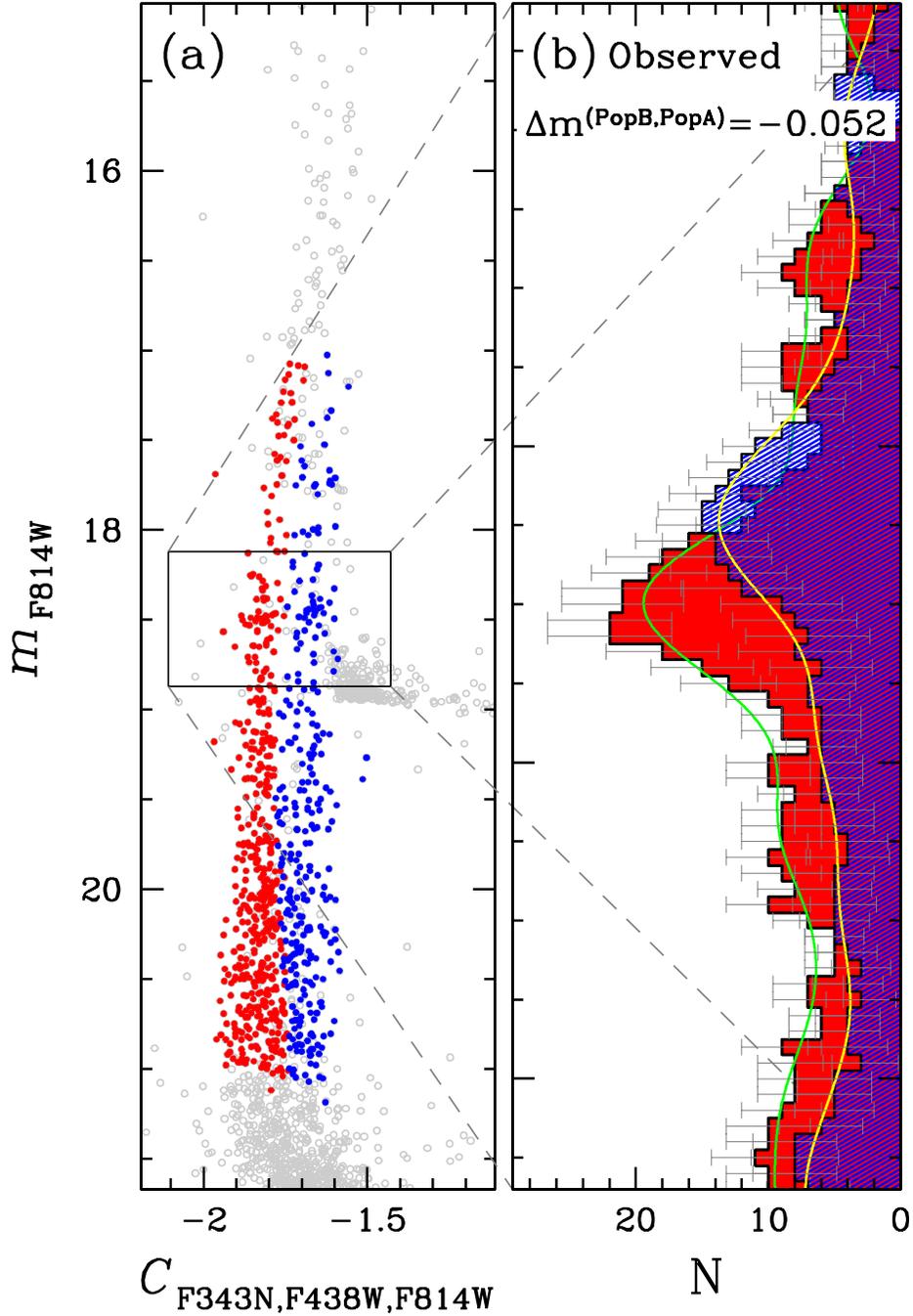}
\caption{\textit{Panel (a)}: $m_{F814W}$ vs. $C_{F343N,F438W,F814W}$ pseudo CMD
of \ngc121. \textit{Panel (b)}: LF of the PopA and PopB stars included in the
black box plotted in the left panel. The gray error bars indicate the
uncertainty relative to the star count of the stars in each histogram bin. The
overplotted green and yellow curves indicate the kernel density estimate of the
PopA and PopB magnitude distribution. The peak of each curve identifies
the RGBB magnitude of the corresponding stellar population.\label{fig:lf121}} 
\end{figure*}

The $m_{F814W}$ vs. $C_{F343N,F438W,F814W}$ CMD of the cluster is shown in the
panel (a) of Figure~\ref{fig:lf121}. The black box, centered on the approximate
location of the PopA and PopB RGBB, is extended over 0.8 mag along the
$m_{F814W}$ axis and includes the stars for which we determined the relative LF,
plotted in panel (b). Both the PopA and PopB LFs, represented by the
corresponding color histogram, were built by dividing the displayed magnitude
range into a grid of $m_{F814W}^k$ points 0.01 mag apart. For each point we
counted the number of stars in the bin $[m_{F814W}^k-0.05,m_{F814W}^k+0.05]$. A
smoother representation of each LF was obtained with a kernel density estimate
of the magnitude distribution by employing a Gaussian kernel with $\sigma =
0.04$ mag. The resulting probability distribution function has been overplotted
as a green curve for the PopA LF and a yellow curve for the PopB LF. The
corresponding maxima mark the location of the PopA and PopB RGBB while their
magnitude difference, $\Delta m^{(PopB,PopA)}$, has been reported on the top of
the panel. 

The uncertainty associated to the $\Delta m^{(PopB,PopA)}$ measurement was
obtained as the sum in quadrature of the standard errors of the PopA and
PopB RGBB magnitude. Each standard error corresponds the 68.27th percentile of
the distribution of the estimates of RGBB magnitude obtained by performing
1000 bootstrapping tests on the corresponding population in the
magnitude interval displayed in panel (b).

We found that the RGBB magnitude separation of \ngc121 in the WFC3 bands is:
$\Delta m_{F336W}^{(PopB,PopA)}$ = $0.020\pm0.020$, $\Delta
m_{F343N}^{(PopB,PopA)}$ = $0.075\pm0.032$, $\Delta m_{F438W}^{(PopB,PopA)}$ =
$-0.069\pm0.024$, $\Delta m_{F814W}^{(PopB,PopA)}$ = $-0.070\pm0.021$; while in
the ACS bands is: $\Delta m_{F555W}^{(PopB,PopA)}$ = $-0.067\pm0.027$, $\Delta
m_{F814W}^{(PopB,PopA)}$ = $-0.052\pm0.017$.

To quantify the significance of a RGBB detection, namely the probability that 
the peaks observed in the LFs are not the result of numerical fluctuations in the
magnitude distribution of stars, we employed a statistical approach, based on the
use of synthetic LFs. The method for the computation of the statistical
significance has been described in detail in \citet{lagioia18}, to which we
refer the interested reader. By using this procedure we found that the statistical
significance of the PopA and PopB RGBBs is higher than 99.3\% in both the ACS
and WFC3 F555W and F814W band.

%%%%%%%%%%%%%%%%%%%%%%%%%%%%%%%%%%%%%%%%%%%%%%%%%%%%%%%%%%%%

\subsection{Comparison with models}
The last step for the determination of the helium variation between PopA and
PopB stars in \ngc121 involves the comparison of the observed $\Delta
m_X^{(PopB,PopA)}$ with the corresponding best-match value obtained
from appropriate theoretical models. 
To this purpose, we took advantage of the same set of isochrones used
for the analysis of synthetic spectra for this cluster (see
Section~\ref{sec:synsp}).

We computed a grid of pairs of isochrones, one with standard helium content ($Y
\approx 0.25$) and the other with enhanced helium content, with a difference in
helium abundance, $\delta Y_i$, ranging from $0.000$ to $0.100$ in steps of
$0.001$.  For each pair of isochrones we simulated two CMDs, each one composed
by 200,000 synthetic stars to account for observational errors in both color and
magnitude. The slope of the LF of the helium standard and helium-enhanced
synthetic CMD was assumed equal to that of the observed PopA and PopB LF,
respectively. For each couple of synthetic CMDs, the estimate of $\Delta m_{i\
X}^{(PopB,PopA)}$ was obtained with the same procedure used for the
observations. As already seen in \citet{lagioia18}, C, N, O and He variation all
contribute to produce the observed magnitude displacement between the two RGBBs.
For this reason we computed, with the method described in
Section~\ref{sec:synsp}, the quantity $\Delta m_{X,CNO}^{(PopB,PopA)}$
($\Delta m_{F555W,CNO}^{(PopB,PopA)}=-0.0059$ mag,
$\Delta m_{F814W,CNO}^{(PopB,PopA)}=-0.0024$ mag), which
indicates the contribution of C, N and O to the
observed magnitude difference, and determined for each simulation, the quantity
$\Delta m_{i\ X,He}^{(PopB,PopA)}$ = $\Delta m_{i\ X}^{(PopB,PopA)} - \Delta
m_{X,CNO}^{(PopB,PopA)}$. Finally we assumed the value of $\delta Y_i$ providing
$\Delta m_{i\ X,He}^{(PopB,PopA)}$ = $\Delta m_X^{(PopB,PopA)} - \Delta
m_{X,CNO}^{(PopB,PopA)}$, as the best estimate of the helium content difference
between PopB and PopA stars, $\delta Y$. We enphasize that we assumed
for PopA and PopB the same metallicity in close analogy with what was done in
similar work by \citet{lagioia18,lee18,milone15c}.

\begin{figure*}
\centering
\includegraphics[width=\textwidth]{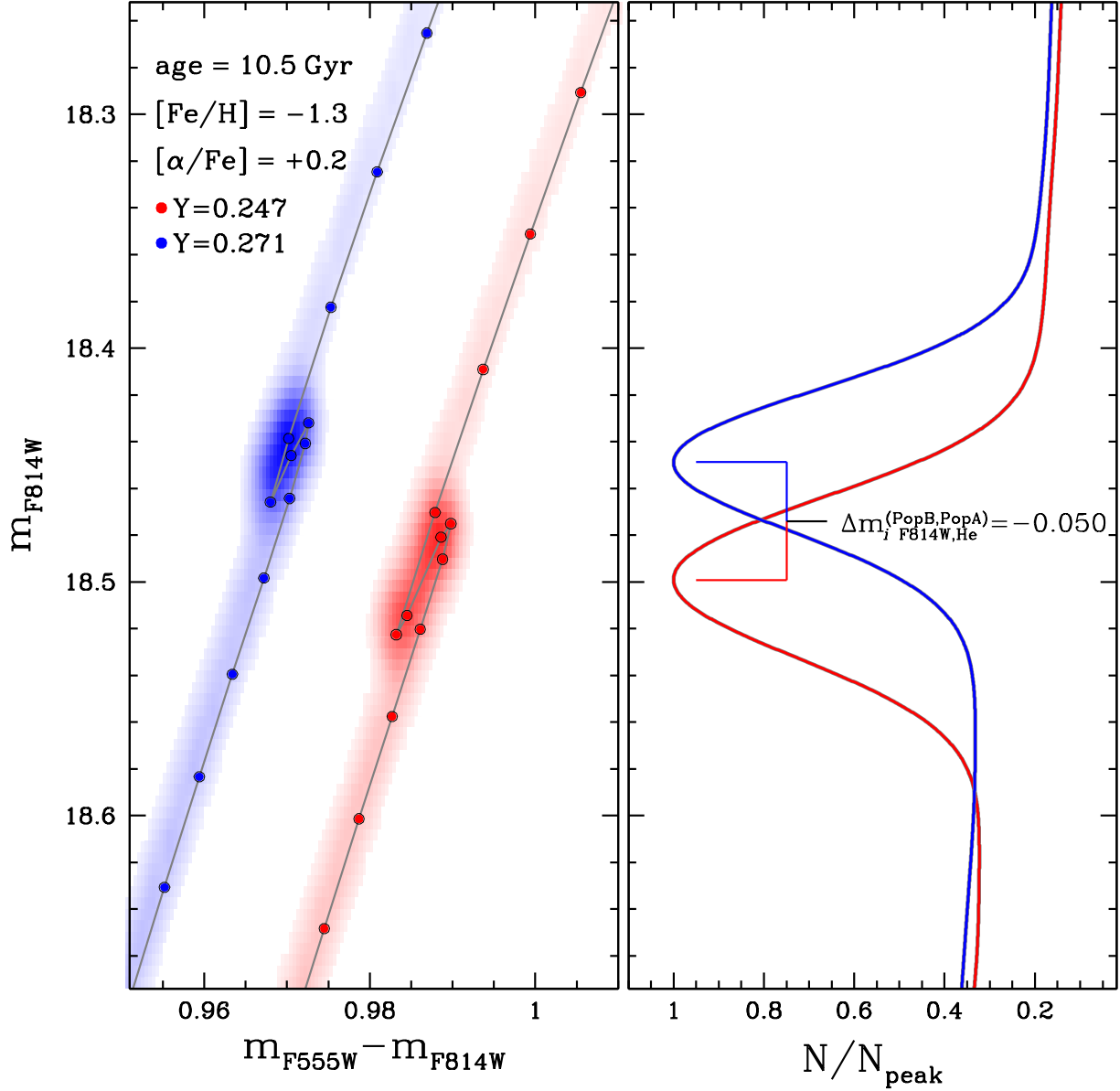}
\caption{\textit{Left panel}: best-fit isochrones for the PopA and PopB stars of
\ngc121. The red and blue points mark, respectively, the location of the model
with $Y=0.247$ and $Y=0.271$ in the $m_{F814W}$ vs $m_{F555W}-m_{F814W}$ CMD.
The red and blue shaded areas represent the Hess diagram of the synthetic CMDs
obtained from the two isochrones. \textit{Right panel}: kernel density estimate
of the magnitude distribution of the two synthetic stellar populations. The
magnitude difference between the two synthetic RGB Bumps is indicated.
\label{fig:dbump}}
\end{figure*}

In Figure~\ref{fig:dbump} we show the procedure for the estimate of $\delta
Y$ in the F814W band for \ngc121. In the left panels two isochrones, the red
with $Y=0.247$ and the blue with $Y=0.271$, corresponding to the best estimate
of $\delta Y_i$ for this cluster, have been overplotted to the corresponding
synthetic CMDs. The helium-enhanced isochrone has been obtained by linearly
interpolating the models with $Y=0.247$ and $Y=0.280$. Age, \feh\ and \afe\ of
the adopted models are reported in the upper-left corner. In the right
panel the corresponding synthetic LFs, normalized to their peak value, are
shown together with the resulting best-matching magnitude difference, $\Delta
m_{i\ F814W,He}^{(PopB,PopA)} = -0.050$.

We found that the corresponding $\delta Y$ is $0.024\pm0.011$.  Similarly, in
F555W band, we found $\delta Y = 0.028\pm0.014$. Therefore, we considered the
weighted mean of the two estimates, $\delta Y = 0.026\pm0.009$, as our best
estimate of the helium content difference between the two populations.

Noticeably the relative helium abundance derived from the analysis of the RGBB
is is consistent with the value of $\delta Y$ inferred from
multiple RGBs at $\sim1.5\,\sigma$ level only.

%%%%%%%%%%%%%%%%%%%%%%%%% FIGURE 10 %%%%%%%%%%%%%%%%%%%%%%%%%%%
%%%%%%%%%%%%%%%%%%%%%%%%%%%%%%%%%%%%%%%%%%%%%%%%%%%%%%%%%%%%

\section{Summary} \label{sec:conclus}
In recent work, we introduced a new method to constrain the relative helium
abundance of MPs in GCs from multi-band photometry of MS, RGB, and RGB bump
stars and find that internal variations of helium are a distinctive feature of
old Galactic GCs \citep[e.g.][]{milone15b}. Specifically, from the study of a
large sample of Galactic GCs, we found that, on average, second-population stars
(2G) are enhanced by $\sim$0.01 in helium mass fraction with respect to the
first population stars \citep[1G, ][]{lagioia18,milone18b}. 

These results provide strong constraints on the physical processes responsible
for the formation of the MPs \citep[e.g.][]{renzini15,dantona16,chantereau17,gieles18}.

In this work we exploited multi-band \hst\ photometry in four SMC GCs, namely
\ngc121, \ngc339, \ngc416 and Lindsay\,1, to investigate, for the first time,
the internal helium variations by using RGB stars. We confirm previous results
by \citet{nieder17a,nieder17b} that the CMDs of these clusters are not
consistent with a simple population and identified the two main populations,
PopA and PopB, along the RGB of each cluster.

From the comparison of the observed colors of the PopA and PopB stars with
synthetic spectra with appropriate chemical composition we find that PopB are on
average enhanced in helium by $\sim0.007$ with respect to PopA stars, similarly
to what is observed in Galactic GCs.  Specifically we obtained $\delta
Y=0.009\pm0.006$ for \ngc121, $\delta Y=0.007\pm0.004$ for \ngc339, $\delta
Y=0.010\pm0.003$ for \ngc416 and $\delta Y=0.000\pm0.004$ for Lindsay\,1.

We also provided estimates for the internal variations in the abundance of
carbon, nitrogen, and oxygen and find that PopB are enhanced in N by
$\sim0.4$ -- $0.6$ dex, and depleted in C and O by $\sim0.1$ -- $0.2$ dex and
$\sim0.0$ -- $0.3$ dex, respectively, with respect to PopA stars. In the case of
\ngc121 we identified the RGB bumps of the two main populations and exploited
their luminosities to infer the relative helium content. We find that PopB
stars are enhanced in helium by $\sim0.026$ with respect to PopA, in agreement
at $\sim1.5\,\sigma$ level with the value inferred from the colors of RGB stars.

A variation of helium content among the stars in \ngc121 was already suggested
by \citet{nieder17a}, who tried to qualitatively reproduce the shape of the
cluster HB by simulating a synthetic population of stars with a spread of
helium mass content of $\Delta Y = 0.025$.     

Noticeably, \citet{hollyhead17} measured the CN and CH band strength from
low-resolution spectroscopy of 16 RGB stars in Lindsay\,1, and concluded that
six stars are enhanced in [N/Fe] by $\sim 0.7$ with respect to the majority of
analyzed stars. All the analyzed stars share almost constant carbon abundance.
Our photometric observations of Lindsay\,1 indicate that PopB stars are
enhanced in nitrogen by $\sim 0.45$ dex with respect to PopA stars thus
providing further evidence of significant nitrogen variations in this cluster.
In contrast, our work suggests that nitrogen anticorrelates with carbon,
similarly to what is observed in Galactic GCs. However, caution must be used
when comparing the results of this paper and those by \citet{hollyhead17}.
Indeed the two groups of stars selected in this paper, namely PopA and
PopB, do not necessarily correspond to the sample of N-rich and N-poor
stars by Hollyhead and collaborators.

Most of the studies on MPs are focused on Galactic GCs older than $\sim 11$ Gyr.
The finding of MPs in Magellanic Clouds GCs with ages between $\sim 6$ and 11
Gyr provide further information on the multiple-population phenomenon. These
clusters, indeed, allow to investigate MPs to early stages of the cluster
formation as well as to understand the impact of a different environments.

Our paper provides evidence that the two main populations of four $\sim 6$ and
11 Gyr old GCs differ in helium by $\sim0.007$, in close analogy with what is
observed in old Galactic GCs. The more helium-rich stars in our analyzed SMC
clusters are also more N-rich and more C-O-poor than stars with primordial
helium. These results are consistent with a scenario where MPs in old Galactic
GCs and younger SMC clusters share similar properties.

\acknowledgments
This work has received funding from the European Research Council
(ERC) under the European Union's Horizon 2020 research innovation
programme (Grant Agreement ERC-StG 2016, No 716082 `GALFOR', PI:
Milone), and the European Union's Horizon 2020 research and innovation
programme under the Marie Sklodowska-Curie (Grant Agreement No 797100,
beneficiary: Marino). APM acknowledges support from MIUR through the
the FARE project R164RM93XW `SEMPLICE'.

\facility{HST (ACS, WFC3)}

\bibliography{ms} 

\end{document}